\documentclass[submission, Proceedings]{SciPost}

\hypersetup{
    colorlinks,
    linkcolor={red!50!black},
    citecolor={blue!50!black},
    urlcolor={blue!80!black}
}

\newcommand{\be}{\begin{equation}}
\newcommand{\ee}{\end{equation}}

\newcommand{\nef}{N_\text{eff}}

\usepackage[bitstream-charter]{mathdesign}
\DeclareSymbolFont{usualmathcal}{OMS}{cmsy}{m}{n}
\DeclareSymbolFontAlphabet{\mathcal}{usualmathcal}

\begin{document}

\begin{center}{\Large \textbf{
Axions in the Early Universe\\
}}\end{center}

\begin{center}
Francesco D'Eramo\textsuperscript{1,2$\star$}
\end{center}
\begin{center}
{\bf 1} Dipartimento di Fisica e Astronomia, Università degli Studi di Padova, \\Via Marzolo 8, 35131 Padova, Italy
\\
{\bf 2} Istituto Nazionale di Fisica Nucleare (INFN), Sezione di Padova, \\Via Marzolo 8, 35131 Padova, Italy 

*francesco.deramo@pd.infn.it
\end{center}

\begin{center}
October 1, 2022
\end{center}

\definecolor{palegray}{gray}{0.95}
\begin{center}
\colorbox{palegray}{
  \begin{tabular}{rr}
  \begin{minipage}{0.1\textwidth}
    \includegraphics[width=30mm]{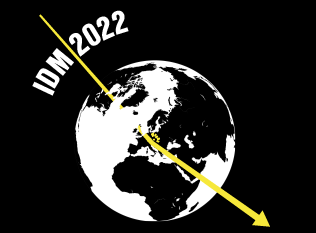}
  \end{minipage}
  &
  \begin{minipage}{0.85\textwidth}
    \begin{center}
    {\it 14th International Conference on Identification of Dark Matter}\\
    {\it Vienna, Austria, 18-22 July 2022} \\
    \doi{10.21468/SciPostPhysProc.?}\\
    \end{center}
  \end{minipage}
\end{tabular}
}
\end{center}

\section*{Abstract}
{\bf
The Peccei-Quinn solution to the strong CP problem provides a motivated framework rich in cosmological consequence. Thermal axion production is unavoidable if there is a thermal bath at early times. Scattering and decay processes of bath particles can dump relativistic axions in the primordial plasma, and they can leave observable signatures in cosmological observables probing both the early and the late universe if produced with a significant abundance. We present recent and significant improvements for the calculation of the axion production rate for different scenarios and apply these results to predict the abundance of produced axions. Finally, we provide updated cosmological bounds on the QCD axion mass.}

\vspace{10pt}
\noindent\rule{\textwidth}{1pt}
\tableofcontents\thispagestyle{fancy}
\noindent\rule{\textwidth}{1pt}
\vspace{10pt}

\section{Introduction}
\label{sec:intro}

The firm observational evidence for some form of dark matter, approximately five times more abundant than ordinary matter, leaves no doubt about the need for physics beyond the standard model. Motivated particle candidates are not postulated to exist ad hoc but they tackle more than one problem. The candidate discussed here, naturally arising from the investigation of symmetries whose violation is small and structured, is one of the best examples. 

The experimentally observed invariance of strong interactions when we flip the arrow of time is rather surprising. The Peccei-Quinn (PQ) mechanism~\cite{Peccei:1977hh,Peccei:1977ur} is an elegant dynamical solution, and all microscopic realizations share some common features. The standard model is extended via a new global symmetry $U(1)_{\rm PQ}$ that must satisfy two key requirements: spontaneously broken, and anomalous under strong interactions. The PQ symmetry breaking scale is bound to be much higher than the energies we can reach with experiments, and the only accessible low-energy residual is a pseudo-Nambu-Goldstone boson (PNGB) arising from the spontaneous PQ breaking. This pseudo-scalar field is known as the QCD axion~\cite{Wilczek:1977pj,Weinberg:1977ma}, and as a consequence of the color anomaly it couples to gluons via the dimension 5 operator
\be
\mathcal{L}_{\rm PQ} \supset \frac{\alpha_s}{8 \pi} \frac{a}{f_a} G^A_{\mu\nu} \widetilde{G}^{A \mu\nu} \ .
\label{eq:LPQaxion}
\ee
Here, we have the QCD fine structure constant $\alpha_s = g_s^2 / (4\pi)$, the gluon field strength tensor $G^a_{\mu\nu}$, and its dual $\widetilde{G}^{a \mu\nu}$. This operator defines the axion decay constant $f_a$. Interactions with other standard model fields are model dependent. Given its PNGB nature, all couplings are suppressed by the scale $f_a$. QCD non-perturbative effects generate a potential once strong interactions confine, and this leads to an axion mass~\cite{GrillidiCortona:2015jxo}
\be
m_a \simeq 5.7 \, \mu eV \, \left( \frac{10^{12} \, {\rm GeV}}{f_a} \right) \ .
\label{eq:ma}
\ee

A light and elusive particle could play a prominent role in the early universe. Famously, the QCD axion is a viable dark matter candidate~\cite{Preskill:1982cy,Abbott:1982af,Dine:1982ah} as discussed in Sec.~\ref{sec:cold}. The PQ framework leads to other fascinating phenomena in the early universe with testable consequences today. A hot axion population, discussed in Sec.~\ref{sec:hot}, is a notable example. We focus on thermal production in Sec.~\ref{sec:thermal}, and we give updated predictions for the KSVZ~\cite{Kim:1979if,Shifman:1979if} and DFSZ~\cite{Zhitnitsky:1980tq,Dine:1981rt} frameworks. We also compare cosmological bounds on flavor-violating axions with laboratory searches and present an updated cosmological bound on the QCD axion mass. Conclusions are given in Sec.~\ref{sec:conclusion}.

\section{Cold Axions}
\label{sec:cold}

The axion contribution to the observed dark matter relic density can be computed by analyzing the field evolution across the expansion history. At early times, at temperatures much higher than the QCD confinement scale, the finite-temperature axion potential is negligible and Hubble friction prevents any motion. The field starts evolving only when the primordial plasma temperature reaches values around the proton mass and the non-perturbative axion potential switches on. The resulting motion is described by damped harmonic oscillations around the minimum of the potential, and the amplitude of such oscillations gets damped by the Hubble expansion as non-relativistic dark matter. This is the misalignment production mechanism.

The resulting energy density today depends on the axion decay constant and the initial field value. For the latter, we adopt the angular variable $\Theta = a / f_a$. An important interplay between the dynamics of PQ breaking and inflationary physics comes into play once one has to specify the initial field value $\Theta_0$. If the PQ phase transition happens after inflation, the two dynamics are decoupled and we can focus on PQ breaking only. The initial axion field value is selected randomly within each causal horizon at the PQ phase transition. Furthermore, when oscillations begin the horizon size is much larger than the one at the PQ phase transition. Thus we have different axion field values within a causal horizon, and one has to average over all possible initial field values $\Theta_0$ that are uniformly distributed between $-\pi$ and $\pi$. This situation has the advantage that the axion relic density depends only on $f_a$. However, the misalignment contribution comes together with the one associated with topological defects formed at the PQ phase transition. Such a network of global axion strings formed at the PQ breaking scale must be evolved down to the nuclear scale. This is a multi-scale problem as a consequence of the large mass hierarchy between the string core tension and the Hubble expansion rate; any numerical approach is extremely challenging. Unfortunately, such a prediction has generated conflicting results for decades, and an agreement is still not reached in the literature~\cite{Gorghetto:2018myk,Gorghetto:2020qws,Buschmann:2019icd,Buschmann:2021sdq,Klaer:2017qhr,Klaer:2017ond}. The current uncertainty is sketched by the red line in Fig.~\cite{Borsanyi:2016ksw}. The knowledge of how such a quantity depends on the axion microscopic parameters is extremely useful since it will allow us to target experimental searches appropriately. On the contrary, if the PQ symmetry is broken during inflation and not restored afterward, the axion field is homogeneous within each causal horizon. Topological defects are inflated away for this case, and the resulting contribution from misalignment depends on both $f_a$ and $\Theta_0$. The blue line in Fig.~\cite{Borsanyi:2016ksw} shows the relation needed to reproduce the relic density.

\begin{figure}
\centering
\includegraphics[width=0.6\textwidth]{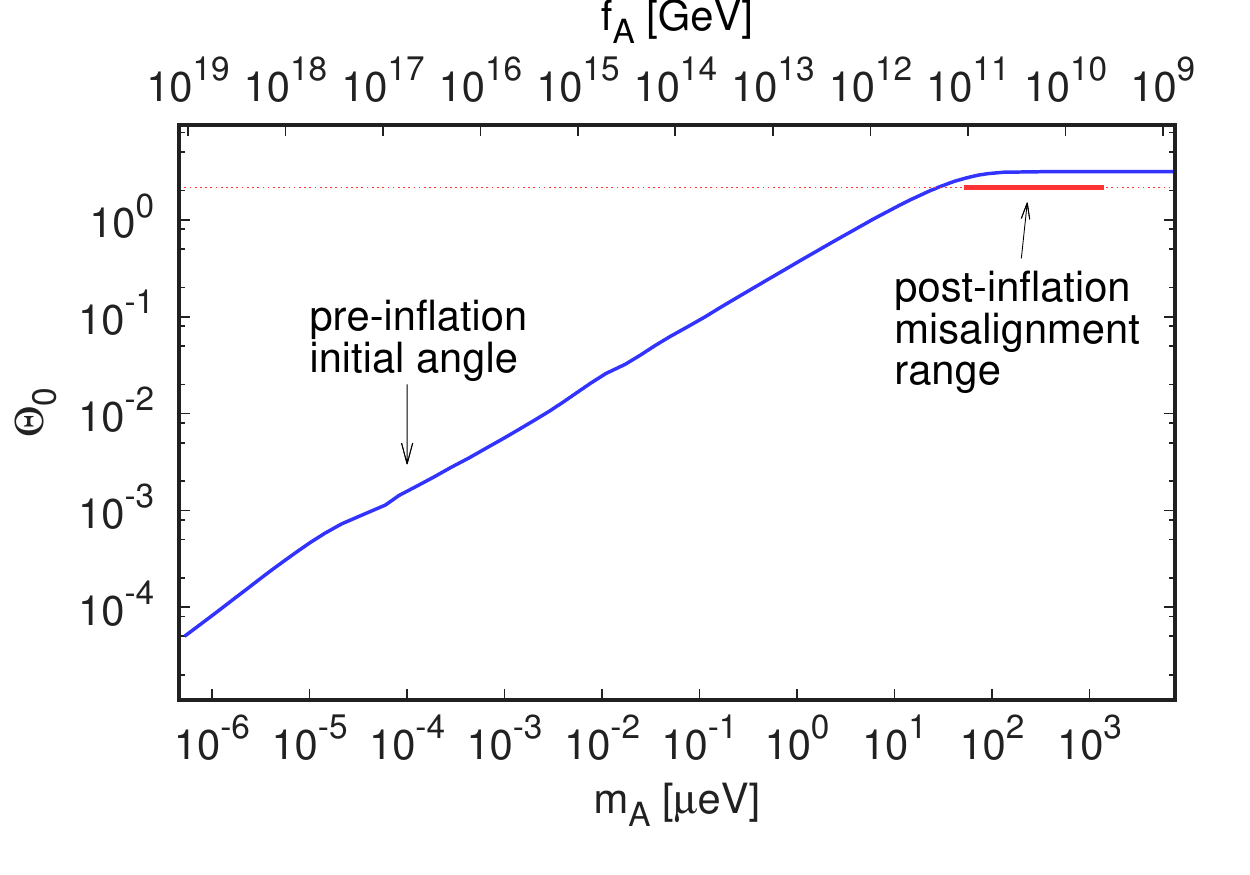}
\caption{Relation between the axion decay constant (or axion mass) and the initial field value to reproduce the observed dark matter density. Figure from Ref.~\cite{Borsanyi:2016ksw}.}
\label{fig:coldaxions}
\end{figure}

\section{Hot Axions}
\label{sec:hot}

Axions can be produced in the early universe with kinetic energy much larger than their rest mass. Several production mechanisms can be responsible for this additional axion population, and in the next section, we will focus on the thermal channels. Regardless of the specific mechanism, their energy gets depleted with the expansion and the overall energy density behaves as radiation as long as these axions remain relativistic. Thus their effect is an additional contribution to the radiation energy density, and we measure this quantity by investigating the formation of light nuclei via Big Bang Nucleosynthesis (BBN) and the formation of the Cosmic Microwave Background (CMB). Historically, this effect has been quantified in terms of a modification of the total number of neutrino species. The effective number of neutrinos $\nef$ is related to the radiation energy density $\rho_\text{rad}$ as
\be
\rho_\text{rad} =  \rho_\gamma \left[ 1+ \frac{7}{8} \left(\frac{T_\nu}{T_\gamma}\right)^{4} \nef \right]\, .
\ee
Any relativistic particle with a substantial energy density, like the axions, will contribute to $\nef$. We look for deviations from the standard cosmological value ($\nef^{\Lambda \text{CDM}}= 3.044$)
\be
\Delta \nef \equiv \nef - \nef^{\Lambda \text{CDM}} = \frac{8}{7} \left(\frac{11}{4}\right)^{4/3} \frac{\rho_a}{\rho_\gamma } \ .
\ee

\section{Thermal QCD Axions}
\label{sec:thermal}

The QCD axion couples to gluons via the anomalous dimension 5 interaction given in Eq.~\eqref{eq:LPQaxion}. Such a coupling is already enough to mediate the thermal production discussed here. Couplings to other standard model particles are model dependent, and they are typically present (perhaps loop-suppressed) unless unnatural cancellations are in action. Broadly speaking, there are two classes of axion interactions with visible matter
\be
\mathcal{L}_{\rm axion-int} \supset  \frac{1}{f_a} \left[  a\, c_X \dfrac{\alpha_X}{8\pi} \, X^{a \mu\nu}\widetilde{X}^a_{\mu\nu} + 
\partial_\mu a \; c_\psi \overline{\psi} \gamma^\mu \psi \right]    \ .
\label{eq:axionINTgeneric}
\ee
The first class contains operators with standard model gauge bosons ($X=\{G,\,W, \,B\}$), and we set $c_G = 1$ consistently with Eq.~\eqref{eq:LPQaxion}. The standard model fermions appearing in the second class are the ones with-defined electroweak gauge quantum numbers ($\psi = \left\{Q_L, u_R, d_R, L_L, e_R \right\}$), and their interactions with the axion preserve the shift symmetry $a \, \rightarrow \, a + {\rm const}$.

Thermal production is, in some sense, an unavoidable injection source since it only assumes the existence of a thermal bath at early times. For every specific model, the amount of axions produced thermally is determined unless we move away from a standard cosmological history. Typically, the leading production channels are binary scatterings and decays. Axions produced via this mechanism are relativistic because these processes are efficient above BBN, and they subsequently red-shift with the expansion until they get to the non-relativistic regime. Such a transition can happen quite late. As a rule of thumb, the axion phase-space distribution stays thermal with a temperature not too much different from the ones of the radiation bath (given their thermal origin). Therefore they manifest themselves as additional radiation at BBN if $m_a \lesssim {\rm MeV}$, and at recombination if $m_a \lesssim 0.3 \, {\rm eV}$.

The tool that we employ to track the hot axion population in the early universe is the Boltzmann equation for the axion number density
\be
\frac{d n_a}{d t} + 3 H n_a = \mathcal{C}_a \ .
\label{eq:BE}
\ee
The left hand simply accounts for the geometry of the expanding universe. The dynamics are on the right-hand side. Thus our goal is to compute the so-called ``collision terms'' that account for all processes changing the number of axions between the initial and final state and solve the resulting Boltzmann equation. After interactions stop happening, the right-hand side vanishes and the comoving axion number density $Y_a = n_a / s$ reaches a constant value $Y_a^\infty$. The resulting contribution to the additional neutrino species results in $\Delta \nef \simeq 74.85 \, (Y^\infty_a )^{4/3}$.

The collision rate can be computed for each one of the operators in Eq.~\eqref{eq:axionINTgeneric}. And one can calculate the number of axions produced if we switch on a single operator at the time~\cite{Ferreira:2018vjj,DEramo:2018vss,Arias-Aragon:2020qtn,Arias-Aragon:2020shv,Green:2021hjh}. However, this situation is not always realistic because once we write down a UV complete model multiple couplings contribute to axion production at different temperatures. Thus we need the production rate across the entire expansion history to quantify axion production for specific UV complete models.

\begin{figure}
\centering
\includegraphics[width=0.45\textwidth]{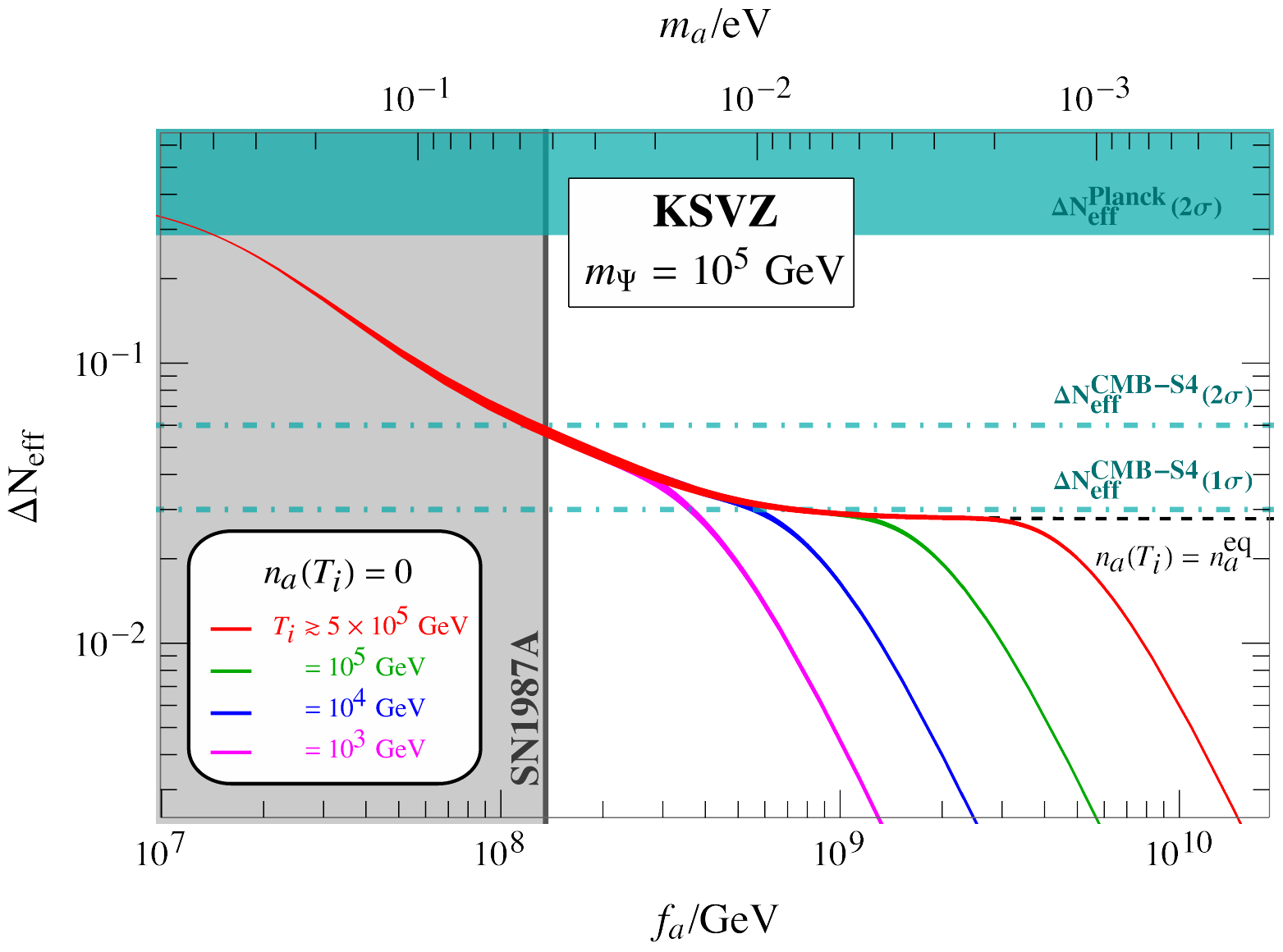} $\qquad\quad$
\includegraphics[width=0.45\textwidth]{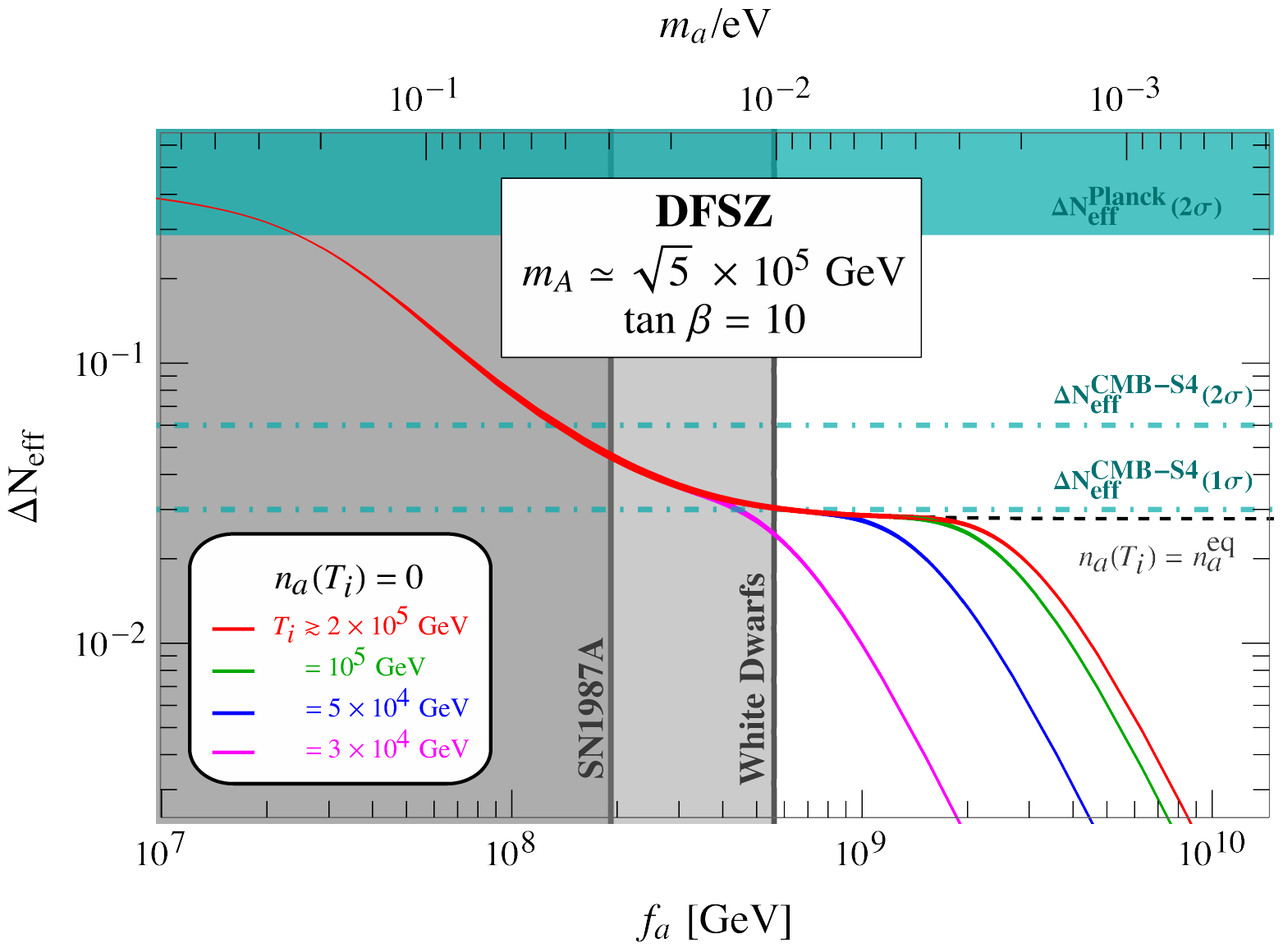}
\caption{$\Delta N_{\rm eff}$ as a function of the axion decay constant for the KSVZ axion (left panel) and the DFSZ axion (right panel). Figures from Ref.~\cite{DEramo:2021lgb}.}
\label{fig:hotaxions}
\end{figure}

The study in Ref.~\cite{DEramo:2021lgb} completed the calculation for the axion production rate at all temperatures for the two most popular classes of axion models: KSVZ~\cite{Kim:1979if,Shifman:1979if} and DFSZ~\cite{Zhitnitsky:1980tq,Dine:1981rt}. For the former, none of the standard model particles transforms under PQ and the color anomaly is due to the presence of new heavy and colored fermions. For the latter, there are no new fermions in the spectrum and the color anomaly is due to standard model quarks. A common feature of both frameworks, and actually of every UV complete model, is the presence of several mass thresholds across which the axion production rate changes its behavior drastically with the temperature. A threshold common to all PQ theories, which is a consequence of the interaction in Eq.~\eqref{eq:LPQaxion} needed to solve the strong CP problem, is the QCD confinement scale. The analysis in Ref.~\cite{DEramo:2021psx} provided a continuous result for the production rate by extending previous calculations above such a scale, and with a smooth interpolation in the between. Another mass threshold present within the KSVZ framework is the one associated with the heavy-colored fermions responsible for the anomaly; the operator in Eq.~\eqref{eq:LPQaxion} is local only well below their masses, and the fermions themselves are dynamical degrees of freedom mediating axion production at higher temperatures.\footnote{The work in Ref.~\cite{Bae:2011jb} discusses the analogous effect for axino production in supersymmetric PQ theories.} For the DFSZ case the situation is ever richer due to the presence of several mass thresholds and the fact that all standard model particles are charged under PQ. First, this case features two Higgs doublets and the mass scale of the heavy Higgs bosons has to be taken into account carefully since axion interactions are super-renormalizable at high temperatures. Furthermore, the electroweak phase transition is another important cosmological phase across which the axion production rate changes its behavior with the temperature significantly as discussed in detail by Ref.~\cite{Arias-Aragon:2020shv}. 

Predictions for the amount of axion dark radiation, expressed in terms of an effective number of additional neutrinos $\Delta N_{\rm eff}$, are shown in Fig.~\ref{fig:hotaxions}. The parameter chosen for the model are specified in the legend, and different lines correspond to different temperatures where the Boltzmann code begins its evolution with vanishing axion population in the early universe. We notice how the region probed currently by Planck is already excluded by supernovae bounds; it is worth keeping in mind that these astrophysical bounds come with their caveats, and it is beneficial to have a complementary probe of that parameter space region. Future CMB-S4 surveys have the potential of probing unexplored regions of the parameter space. This is particularly true for the KSVZ axion illustrated in the left panel of the figure. For the DFSZ axion, shown in the right panel, the CMB-S4 discovery reach is around the same region as the one for white dwarf bounds (present in this case because the DFSZ axion couples to the electron). 

The methodology adopted so far can be employed to provide predictions also for other UV completions. A broad class of theories widely explored in the recent literature is the one where axion couplings to standard model fermions are non-diagonal. The general Lagrangian capturing all models where this happens reads
\be
\mathcal{L}^{(a)}_{{\rm FV}} = \frac{\partial_\mu a}{2 f_a} \sum_{\psi_i \neq \psi_j} \, \bar{\psi}_i \gamma^\mu \left( c^V_{\psi_i \psi_j} + c^A_{\psi_i \psi_j} \gamma^5 \right) \psi_j \ .
\label{eq:LaFV}
\ee
These flavor-violating couplings arise in theories connecting the origin of the SM flavor structure with the PQ symmetry. Even if the high-scale theory preserves the flavor symmetry, they can arise from quantum corrections due to flavor-violating standard model interactions. They are the target of several terrestrial experiments searching for rare meson decays. The early universe offers a complementary probe for this class of theories as well. The analysis by Ref.~\cite{DEramo:2021usm} performed a systematic study of these flavor-violating couplings by switching one operator at a time. For the first time, the contribution from binary scatterings was included and it was shown how it dominates completely the total axion production rate for the case of quarks. Scatterings were previously overlooked in the literature because decays always dominate unless there are hierarchies among the couplings mediating these processes. However, we can attach a QCD/QED interaction vertex to the three-point interactions in Eq.~\eqref{eq:LaFV} and find a contribution  proportional to the strong coupling constant that is significantly enhanced for quarks. Furthermore, the treatment of the QCD crossover was improved via techniques analogous to the ones adopted for the flavor-conserving case.

\begin{figure}
\centering
\includegraphics[width=0.45\textwidth]{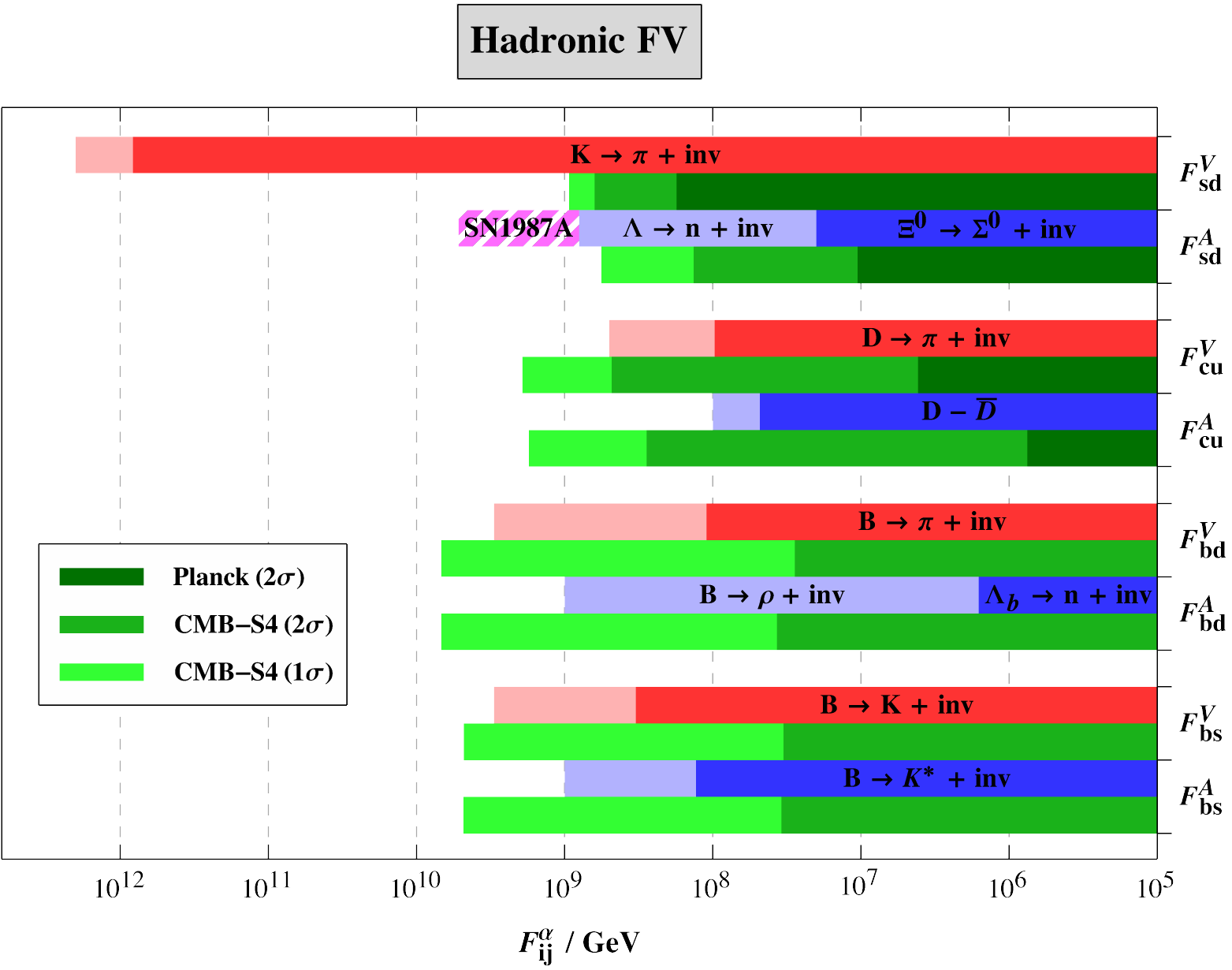} $\qquad\quad$
\includegraphics[width=0.45\textwidth]{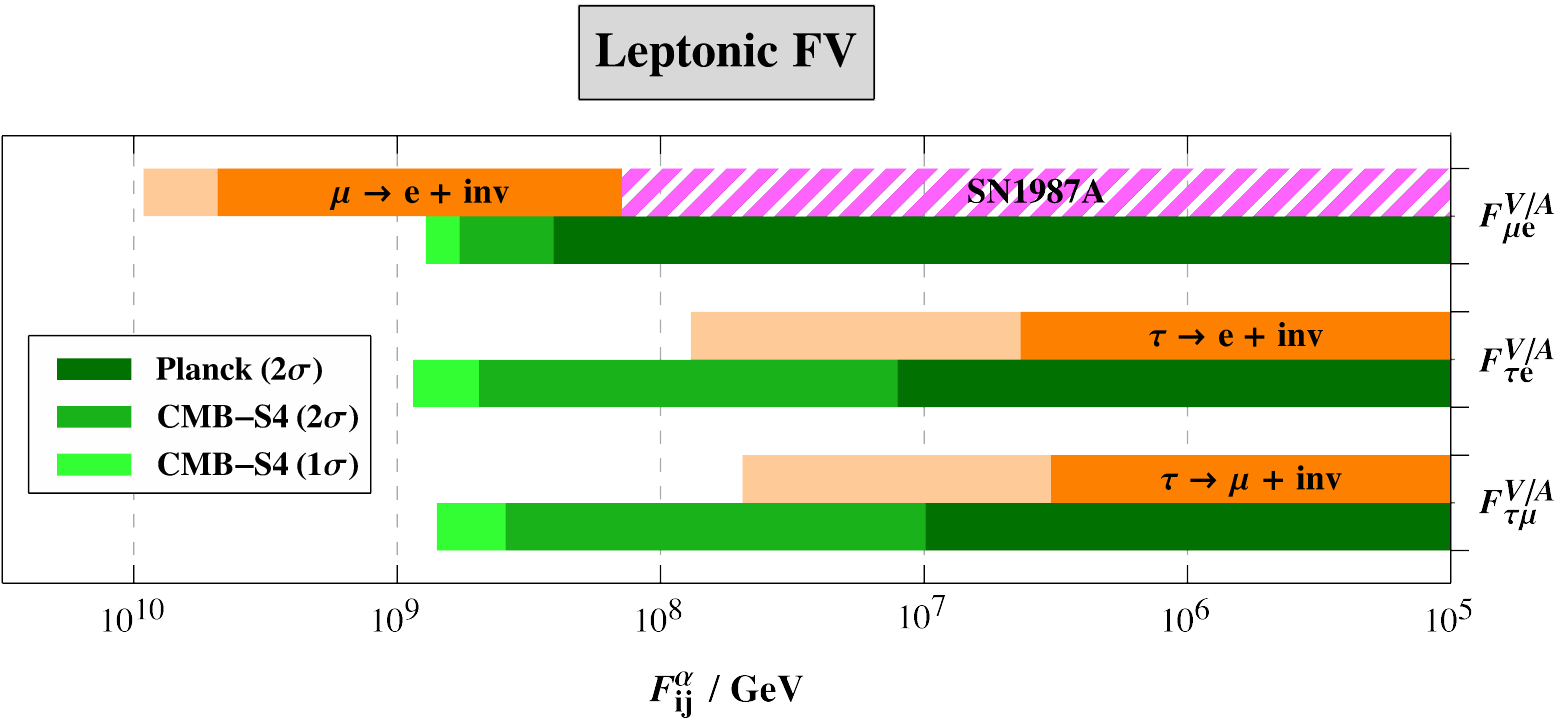}
\caption{Current bounds and future prospects for flavor-violating axion interactions with quarks (left panel) and leptons (right panel). Figures from Ref.~\cite{DEramo:2021usm}.}
\label{fig:FVaxions}
\end{figure}

The results in Fig.~\ref{fig:FVaxions} show the comparison between terrestrial and cosmological bounds for flavor-violating axion interactions with quarks and leptons. Dark bands are current bounds, faint bands are projections for the future. The green bands (both dark and faint) show current and future cosmological bounds. For light quarks and leptons, we notice how laboratory searches perform better than cosmology. The situation changes once we include the third fermion generation where cosmology is competitive with terrestrial experiments and, in some cases, achieves better bounds. 

\begin{figure}
\centering
\includegraphics[width=0.45\textwidth]{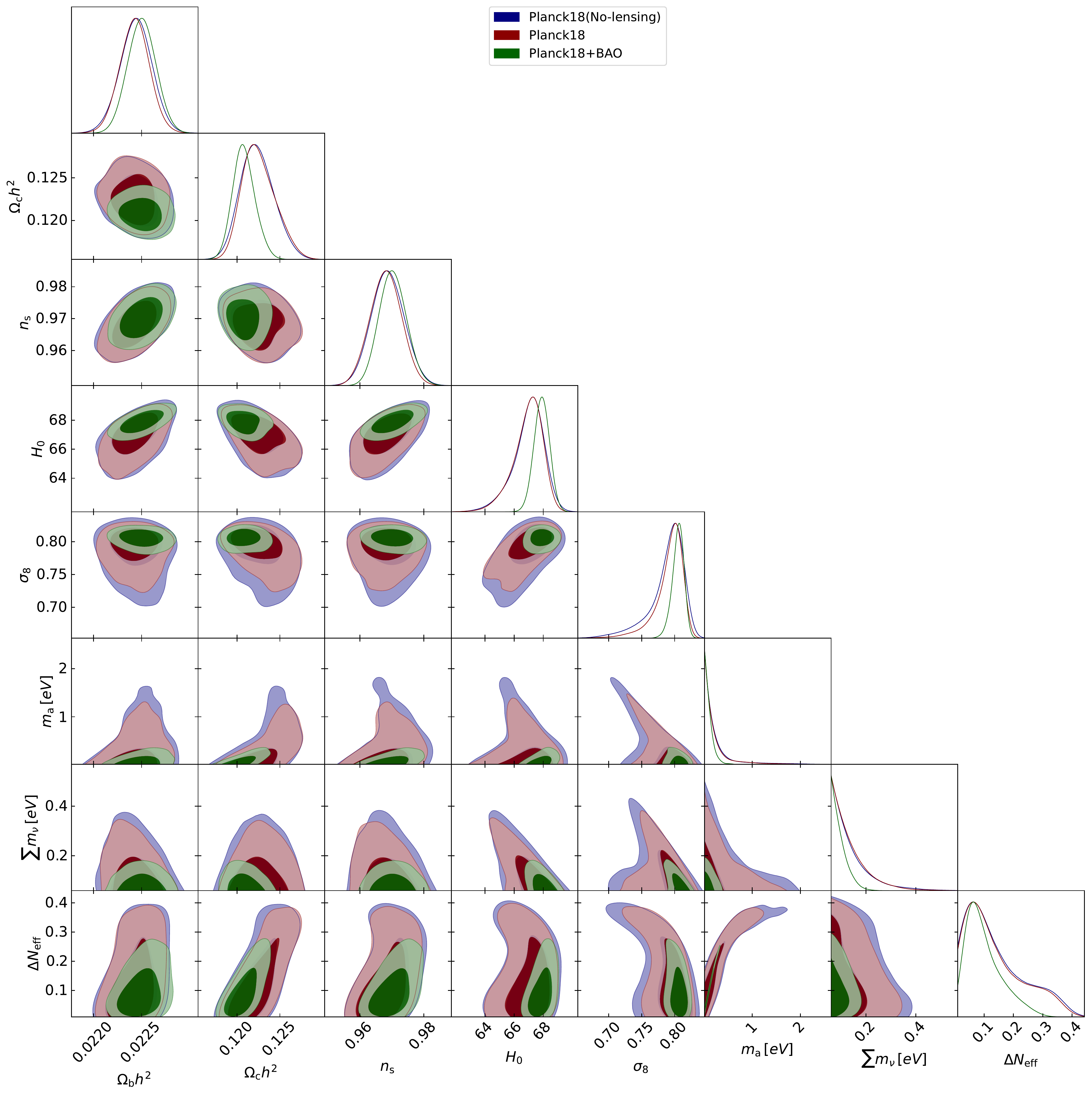} $\qquad\quad$
\includegraphics[width=0.45\textwidth]{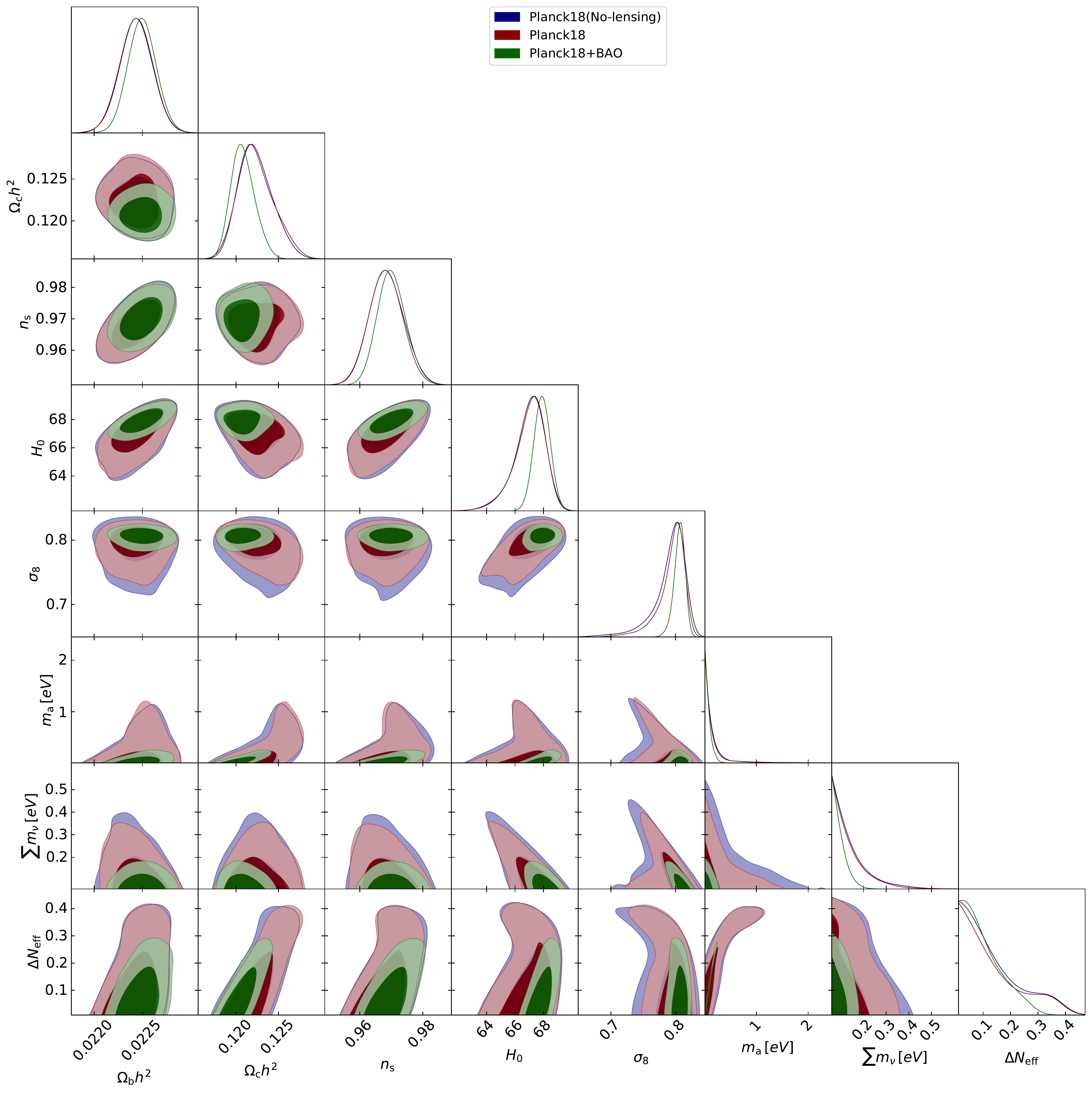}
\caption{Allowed regions and one-dimensional probability posterior distributions for the KSVZ axion (left panel) and the DFSZ axion (right panel). Figures from Ref.~\cite{DEramo:2022nvb}.}
\label{fig:cosmo}
\end{figure}

All the results presented so far neglected the axion mass. While this is definitely allowed at the time of BBN, it is not always the case at the time of recombination if the expression in Eq.~\eqref{eq:ma} approaches values around $0.1 \, {\rm eV}$. The recent analysis by Ref.~\cite{DEramo:2022nvb} incorporated a finite axion mass consistently for the KSVZ and DFSZ scenarios. The plots in Fig.~\ref{fig:cosmo} show the outcome of such a complete cosmological analysis that included also baryon acoustic oscillations (BAO). We report here the updated cosmological bounds on the QCD axion mass
\be
m_a \leq \left\{ 
\begin{array}{ccc}
0.282 \, (0.402) \; {\rm eV} & $\qquad$ & \text{KSZV axion} \\
0.209 \, (0.293) \; {\rm eV} & $\qquad$ & \text{DFSZ axion} 
\end{array} 
\right. \ .
\ee
Here, the results are shown at $95\%$ and $99\%$ CL, and the bound is stronger for the DFSZ axion because it couples to all standard model particles and more production channels contribute. These bounds, obtained with the production rates of Ref.~\cite{DEramo:2021lgb}, are about a factor of 5 stronger than the ones found by previous analysis that adopted approximate results for the rates.

\section{Conclusions}
\label{sec:conclusion}

The PQ framework is one of the strongest motivated particle scenarios for physics beyond the standard model. The low-energy residual, which is known as the QCD axion, kills two birds with one stone: the energy stored in the field oscillations behaves like cold dark matter. Axion detection is rather difficult because axion couplings are suppressed by the large PQ breaking scale. Notwithstanding these challenges, several collaborations have been pushing the high-intensity frontier of particle physics by introducing novel detection strategies and putting forward concrete proposals for experiments. Besides searches in terrestrial laboratories, axion interactions lead to a plethora of rich phenomena in the early universe that leave an imprint in cosmological observables today. The production of cold dark matter is the reason why the QCD axion is one of the strongest motivated candidates from the top down, but it is far from being the only cosmological consequence of the PQ framework. 

Thermal production of relativistic axions in the early universe is unavoidable if we have a thermal bath. This hot axion population leaves an imprint in the CMB as shown in Fig.~\ref{fig:hotaxions} for two broad classes of axion UV complete models. We point out also an intriguing interplay with flavor physics that we have just started to explore in Fig.~\ref{fig:FVaxions}. The methodology presented here is general and it can be employed to analyze other microscopic models such as explicit scenarios of flavor-violating couplings.

\newpage

\section*{Acknowledgements}

\paragraph{Funding information} The research reported here is supported by the following research grants: ``The Dark Universe: A Synergic Multi-messenger Approach'' number 2017X7X85K under the program PRIN 2017 funded by the Ministero dell'Istruzione, Universit\`a e della Ricerca (MIUR); ``New Theoretical Tools for Axion Cosmology'' under the Supporting TAlent in ReSearch@University of Padova (STARS@UNIPD). The author is supported by Istituto Nazionale di Fisica Nucleare (INFN) through the Theoretical Astroparticle Physics (TAsP) project, and by the European Union's Horizon 2020 research and innovation programme under the Marie Sk\l odowska-Curie grant agreement No 860881-HIDDeN. 

\bibliography{DEramo_Axions.bib}

\nolinenumbers

\end{document}